\documentstyle[12pt,epsf]{article}
\newcommand{\sect}[1]{ \section{#1} \setcounter{equation}{0} }

\newcommand{\kslash}{k \! \! \! /}

\newcommand{\half}{\mbox{\small{$\frac{1}{2}$}}}

\newcommand{\Nf}{N_{\!f}} 
\newcommand{\Nff}{\tilde{N}_{\!f}} 
 
\begin{document}
\title{Large $\Nf$ calculations in deep inelastic scattering}
\author{J.F. Bennett\footnote{Talk presented at Quarks '98, Suzdal, Russia.}\,
\& J.A. Gracey, \\ Theoretical Physics Division, \\ 
Department of Mathematical Sciences, \\ University of Liverpool, \\ 
Peach Street, \\ Liverpool, \\ L69 7ZF, \\ United Kingdom.}
\date{} 
\maketitle 
\vspace{5cm}

\noindent 
{\bf Abstract.} We describe the evaluation of the anomalous dimensions of 
twist-2 deep inelastic light cone operators to $O(1/\Nf)$ as a check on future 
perturbative calculations. In particular we present recent results for the 
singlet gluonic operator dimension in polarised and unpolarised scattering and 
give three loop predictions for the $O(1/\Nf)$ gluonic eigenoperator. The 
section of the 3-loop $O(1/\Nf)$ DGLAP splitting function proportional to the 
adjoint quadratic Casimir is also calculated for the singlet gluonic operators.

\vspace{-19.5cm} 
\hspace{13.4cm} 
{\bf LTH-432} 

\newpage

\sect{Introduction.} 

Our modern view of hadronic matter is based upon the rich quantum field theory 
of Quantum Chromodynamics (QCD) (for a review see \cite{1}) in which we picture
hadrons as bound states of valence quarks in a sea of vacuum excitation 
quark/antiquark pairs and gluons. Because QCD has the remarkable property of 
asymptotic freedom \cite{2} we can use high momentum transfer reactions such as
deep inelastic scattering $( lepton$ $+$ $nucleon$ $\longrightarrow$ $lepton$ 
$+$ $hadrons )$ to compare experiment with the predictions of perturbative QCD 
(pQCD) and so refine our ideas on the structure of strongly interacting 
particles.

The formal QCD approach to deep inelastic scattering (DIS) allows us to predict
structure function behaviour using moment sum rules. Key to this method is the 
formalism of the light cone expansion (LCE) in which the non-local time ordered
product of electromagnetic quark currents appearing in the general DIS 
cross-section is expanded in a series of local, spin-$n$ operators together 
with $c$-number Wilson co-efficients. The LCE is easily seen to be dominated by
operators of lowest twist $( \tau$ $=$ $operator$ $dimension$ $-$ $operator$ 
$spin )$. For QCD the lowest twist operators available are twist-$2$. It can be
shown by using the anomalous dimensions associated with the renormalisation of 
insertions of these operators in 2-point Green functions, that we can determine
the evolution of the moment sum rules with the large momentum transfer scale.

The unpolarised twist-2 operators are \cite{3},
\begin{eqnarray}
{\cal O}^{{\mbox{\footnotesize{NS}} i}}_{\mu_1 \ldots \mu_n} &=&
i^{n-1} {\cal S} \bar{\psi} \gamma_{\mu_1} D_{\mu_2} \ldots D_{\mu_n}
\half \lambda^{i} \psi \nonumber \\
{\cal O}^{{\mbox{\footnotesize{q}}}}_{\mu_1 \ldots \mu_n} &=&
i^{n-1} {\cal S} \bar{\psi} \gamma_{\mu_1} D_{\mu_2} \ldots D_{\mu_n} \psi
\nonumber \\
{\cal O}^{{\mbox{\footnotesize{g}}}}_{\mu_1 \ldots \mu_n} &=&
\half i^{n-2} {\cal S}~G_{\mu_1\alpha} D_{\mu_2} \ldots D_{\mu_{n-1}}
G^{\alpha}_{~\,\mu_n}
\label{unpol}
\end{eqnarray}
and the polarised twist-2 operators are \cite{4},
\begin{eqnarray}
{\cal R}^{{\mbox{\footnotesize{NS}} i}}_{\sigma\mu_1 \ldots \mu_{n-1}} &=&
i^{n-1} {\cal S} \bar{\psi} \gamma_5 \gamma_{\sigma} D_{\mu_1} \ldots 
D_{\mu_{n-1}} \half \lambda^{i} \psi \nonumber \\
{\cal R}^{{\mbox{\footnotesize{q}}}}_{\sigma\mu_1 \ldots \mu_{n-1}} &=&
i^{n-1} {\cal S} \bar{\psi} \gamma_5 \gamma_{\sigma} D_{\mu_1} \ldots 
D_{\mu_{n-1}} \psi \nonumber \\
{\cal R}^{{\mbox{\footnotesize{g}}}}_{\sigma\mu_1 \ldots \mu_{n-1}} &=&
\half i^{n-2} {\cal S}~\epsilon_{\sigma\alpha\beta\gamma} G^{\beta\gamma} 
D_{\mu_1} \ldots D_{\mu_{n-2}} G^{\alpha}_{~\, \mu_{n-1}}
\label{pol}
\end{eqnarray}
Here, $\psi$ is the quark field, $G_{\mu\nu}$ is the gluon field strength 
tensor, $D_{\mu}$ is the QCD covariant derivative, $\lambda^{i}$ are the 
generators of $SU(\Nf)$, ${\cal S}$ stands for symmetrisation over Lorentz 
indices and \mbox{NS} denotes the non-singlet operator under $SU(\Nf)$ as 
opposed to the other singlet operators.

An alternative, less abstract means of providing theoretical predictions for 
DIS is given by the parton model. Here the structure functions may be expressed
in terms of parton distribution functions (PDFs). These give the probability 
that the struck nucleon constituent (parton) carries a particular fraction of 
the total nucleon momentum. The evolution of the PDFs are then governed by the 
DGLAP equation \cite{5}. This contains the DGLAP `splitting functions' which 
express the probabilities for the struck parton undergoing certain collinear 
decays as the energy scale changes. The DGLAP splitting functions can be 
explicitly obtained from the previously mentioned operator anomalous dimensions
through an inverse Mellin transform. Solutions of the full DGLAP equation 
provide us with scale dependent PDFs which may than be used as inputs for other
hard scattering processes where the formal QCD approach is not necessarily 
applicable as well as giving us a deeper insight into hadron structure in the 
asymptotic regime. 

With facilities such as HERA currently opening up new kinematical regimes for 
unpolarised and (soon) polarised DIS, there presently exists a need to perform
a NNLO twist-2 analysis to ensure a high precision examination of pQCD. This is
especially important where the new order's effects are expected to be seen 
experimentally (as in low $x$, high $Q^{2}$ scattering at HERA). This programme
requires the 2-loop Wilson co-efficients and 3-loop anomalous dimensions of all
the DIS twist-2 light cone operators with full spin dependence. Although the 
complete 2-loop operator dimensions have been known for many years now 
\cite{3,6,7,8}, the full 3-loop computations have only become viable in recent 
years. The current state of play is that the 3-loop unpolarised singlet and 
non-singlet operator dimensions have been calculated for particular spins of 
the operator ($n$ $=$ $2$, $4$, $6$, $8$ and $n$ $=$ $2$, $4$, $6$, $8$, $10$ 
respectively) \cite{9,10}. In addition, the 2-loop finite parts required for 
the full 3-loop results have recently been calculated for all twist-$2$ 
unpolarised and polarised operators \cite{11,12}. As one might expect, these 
are particularly difficult calculations involving thousands of Feynman 
diagrams. It is clear that a check on the final full $3$-loop results would be 
useful. One way of approaching this would be to calculate the dimensions using 
an alternative expansion parameter to the QCD coupling and check that there is 
agreement where overlap exists. We can do this by use of the $1/\Nf$ expansion 
and a critical point approach.

\sect{The Method.} 

By applying the critical large $\Nf$ method developed in the series of papers 
\cite{13} it is possible to obtain expressions for the twist-2 operator 
dimensions to\, $O(1/\Nf)$. The method entails the analysis of operator 
insertions in QCD Green functions at a $d$-dimensional non-trivial 
renormalisation group fixed point. This fixed point may be found as a stable 
zero of the $d$-dimensional four-loop QCD $\beta$-function \cite{14} and is 
located at
\begin{eqnarray}
a_c &=& \frac{3\epsilon}{4T(R)\Nf} ~+~ \left[ \ \frac{33}{16}C_2(G) \epsilon
{}~-~ \left( \frac{27}{16}C_2(R) + \frac{45}{16}C_2(G) \right) \epsilon^2 
\right. \nonumber \\
&& +~ \left( \frac{99}{64}C_2(R) + \frac{237}{128}C_2(G) \right) 
\epsilon^3 \nonumber \\
&& \left. +~ \left( \frac{77}{64}C_2(R) + \frac{53}{128}C_2(G) \right) 
\epsilon^4 ~+~ O(\epsilon^5) \ \right] \frac{1}{T^2(R)\Nf^2} ~+~ 
O \left( \frac{1}{\Nf^3} \right) 
\end{eqnarray}
where $a_c$ is the strong coupling constant at criticality, $C_2(R)$ and 
$C_2(G)$ are the fundamental and adjoint quadratic Casimirs respectively, 
$\mbox{tr}(T^aT^b)$ $=$ $\half\delta^{ab}$ for $T^a$ the generators of 
$SU(N_c)$ and $d$ $=$ $4$ $-$ $2\epsilon$.

Several good things come out of using such an approach;

\begin{enumerate}
\item Propagators take on a simple, dressed, scaling form due to the scaling properties of Green functions at a renormalisation group fixed point. 

For example, the quark and gluon propagators are respectively \cite{15},
\begin{equation} 
\psi(k) ~ \sim ~ \frac{A\kslash}{(k^2)^{\mu-\alpha}} ~~~,~~~ 
A_{\mu\nu}(k) ~ \sim ~ \frac{B}{(k^2)^{\mu-\beta}} \left[ \eta_{\mu\nu} ~-~ 
(1-b) \frac{k_\mu k_\nu}{k^2} \right] 
\label{props} 
\end{equation}  
where $b$ is the covariant gauge parameter, $d$ $=$ $2\mu$, $A$ and $B$ are 
momentum independent amplitudes and the exponents $\alpha$ and $\beta$ are 
defined using simple dimensional analysis of the massless QCD action as  
\begin{equation} 
\alpha ~=~ \mu ~-~ 1 ~+~ \half \eta ~~~,~~~ \beta ~=~ 1 ~-~ \eta ~-~ \chi 
\label{defexp} 
\end{equation} 

Here, the critical exponent $\eta$ is the quark anomalous dimension associated 
with the quark wave function renormalisation at the critical coupling and 
similarly $\chi$ is the anomalous dimension of the quark-gluon vertex.

\item Using the critical renormalisation group it can be shown that the form of
a particular $O(1/\Nf)$ operator anomalous dimension at $a_c$ may be calculated
by evaluating the residues of first order poles (we use an analytic 
regularisation of the gluon dimension) in appropriate $O(1/\Nf)$ two-point
Green functions at $a_c$. These Green functions consist of an insertion of the
operator into a QCD two-point function. (We also need to take the field 
renormalisations into account).  

\item Since, in the language of statistical mechanics, we are working at a 
critical point, we know that we have an essentially massless theory which 
enables us to use integration tricks such as uniqueness and conformal
transformations to evaluate diagrams.

\item In $d$-dimensions and to $O(1/\Nf)$ there exists a universality 
equivalence between QCD and the non-abelian Thirring model (NATM) at $a_c$ 
\cite{16}. This may be seen through the reproduction of the QCD triple and 
quartic gluon vertices by fermion loop integration in three and four point NATM
Green functions at $a_c$ using the above propagators. The upshot of this is 
that we can use the simpler NATM interactions in our calculations and dispense 
with the tricky three and four-point gluon vertices.

\item The fact that we work with a fixed $d$-dimensional spacetime with an
analytic rather than a dimensional regularisation seems to alleviate some of
the problems caused when treating $\gamma_5$ in arbitrary dimensions \cite{17}.
This obviously becomes important in the calculation of the dimensions of the 
polarised operators.
\end{enumerate}

\sect{Comparing singlet operator dimensions at $O(1/\Nf)$ with perturbation 
theory}

A slight complication in this calculation arises due to mixing of the singlet 
operators in both polarised and unpolarised scattering. Since the singlet light
cone operators share the same quantum numbers and have equal canonical 
dimension in strictly four dimensions, they mix under renormalisation. This 
means that we have to consider a matrix of renormalisation constants for these 
operators
\begin{equation} 
{\cal O}_{\footnotesize{\mbox{ren}}}^i ~=~ Z^{ij} 
{\cal O}_{\footnotesize{\mbox{bare}}}^j 
\end{equation} 
Here the indices $i,j$ $=$ $q,g$ and we are refering to the operators 
(\ref{unpol}) and (\ref{pol}).  The anomalous dimensions, $\gamma_{ij}(a)$, are
defined by 
\begin{equation} 
\gamma_{ij}(a) ~=~ \left( 
\begin{array}{ll} 
\gamma_{qq}(a) & \gamma_{gq}(a) \\ 
\gamma_{qg}(a) & \gamma_{gg}(a) \\ 
\end{array} 
\right) 
\label{mixmat} 
\end{equation} 
where $\gamma_{ij}(a)$ $=$ $\beta(a) (\partial/\partial a) \ln Z_{ij}$ and 
$\beta(a)$ is the renormalization group function governing the running
of the QCD coupling constant $a$. The entries in $\gamma_{ij}(a)$ depend on the
colour group parameters, $\Nf$ and $n$ and since it is the $1/\Nf$ corrections 
that we are interested in, we define the explicit form of the entries as 
\begin{eqnarray}  
\gamma_{qq}(a) &=& a_1a + (a_{21}\Nff + a_{22})a^2 + (a_{31}\Nff^2 + a_{32}\Nff 
+ a_{33})a^3 + O(a^4) \nonumber \\ 
\gamma_{gq}(a) &=& b_1a + (b_{21}\Nff + b_{22})a^2 + (b_{31}\Nff^2 + b_{32}\Nff 
+ b_{33})a^3 + O(a^4) \nonumber \\ 
\gamma_{qg}(a) &=& c_1 \Nff a + c_2\Nff a^2 + (c_{31}\Nff^2 + c_{32}\Nff 
+ c_{33})a^3 + O(a^4) \nonumber \\ 
\gamma_{gg}(a) &=& (d_{11}\Nff + d_{12})a + (d_{21}\Nff + d_{22})a^2 
+ (d_{31}\Nff^2 + d_{32}\Nff + d_{33})a^3 + O(a^4)  
\label{matdef}
\end{eqnarray} 
where $\Nff$ $=$ $T(R)\Nf$ and the coefficients $a_{ij}$, $b_{ij}$, $c_{ij}$ 
and $d_{ij}$ depend on $n$ and the colour group Casimirs. 

We note that this matrix of anomalous dimensions has eigenvalues
\begin{equation} 
\lambda_\pm(a) ~=~ \frac{1}{2} ( \gamma_{qq} + \gamma_{gg} ) ~\pm~ 
\frac{1}{2} \left[ ( \gamma_{qq} - \gamma_{gg} )^2 + 4 \gamma_{qg}\gamma_{gq} 
\right]^{\half} 
\end{equation}  
Expanding in powers of $a$ and retaining the same orders in $1/\Nf$ with the 
definitions (\ref{matdef}) we find, 
\begin{eqnarray} 
\lambda_-(a) &=& \left( a_1 - \frac{b_1c_1}{d_{11}}\right) a + 
\left( a_{21} - \frac{b_{21}c_1}{d_{11}}\right) \Nf  a^2 + 
\left( a_{31} - \frac{b_{31}c_1}{d_{11}}\right) \Nf^2 a^3 + O(\Nf^3 a^4) 
\nonumber \\ 
\lambda_+(a) &=& \left( d_{11}\Nf + d_{12} + \frac{b_1c_1}{d_{11}} \right) a
+ \left( d_{21} + \frac{b_{21}c_1}{d_{11}} \right) \Nf a^2 \nonumber \\ 
&& ~ + \left( d_{31} + \frac{b_{31}c_1}{d_{11}} \right) \Nf^2 a^3 
+ O(\Nf^3 a^4) 
\label{blobby}
\end{eqnarray} 
Here we can see that $\lambda_+(a)$, $\lambda_-(a)$ are dominated by
contributions from the gluonic and fermionic operators respectively.

When in our approach we consider the singlet sector operators in $d$-dimensions
at $a_c$, we find that they no longer mix. (It is easy to see that the 
difference is $O(\epsilon)$.) This means that by calculating in $d$-dimensions,
we are accessing the above mixing matrix eigenvalues in perturbation theory. 
This is borne out by explicit calculations. By evaluating the graphs required 
for $\gamma_{gg}(a_c)$, $\gamma_{qq}(a_c)$ we actually obtain $\lambda_+(a_c)$ 
and $\lambda_-(a_c)$ respectively, with the universality equivalence between 
QCD and NATM at $O(1/\Nf)$ accounting for the contributions from the 
off-diagonal dimensions.

\sect{Results}

The unpolarised non-singlet operator dimension at $O(1/\Nf)$ was published in 
\cite{18}. The polarised non-singlet operator result was published in \cite{19}
together with the expressions for $\lambda_-(a_c)$ to $O(1/\Nf)$ and 
$\lambda_+(a_c)$ to $O(1)$ for polarised and unpolarised scattering. 

We now give new results \cite{20,21} for $\lambda_+(a_c)$ to $O(1/\Nf)$ for the
unpolarised and polarised gluonic operator together with the $3$-loop 
$n$-dependent eigenvalue predictions for comparison with perturbation theory 
(\ref{blobby}). The relevant graphs are given in Fig. 1 with a different
operator insertion Feynman rule \cite{7,22} used for the separate unpolarised 
and polarised cases.

\vspace{1cm}
\hspace{3cm}{\epsfysize=5cm
\epsfbox{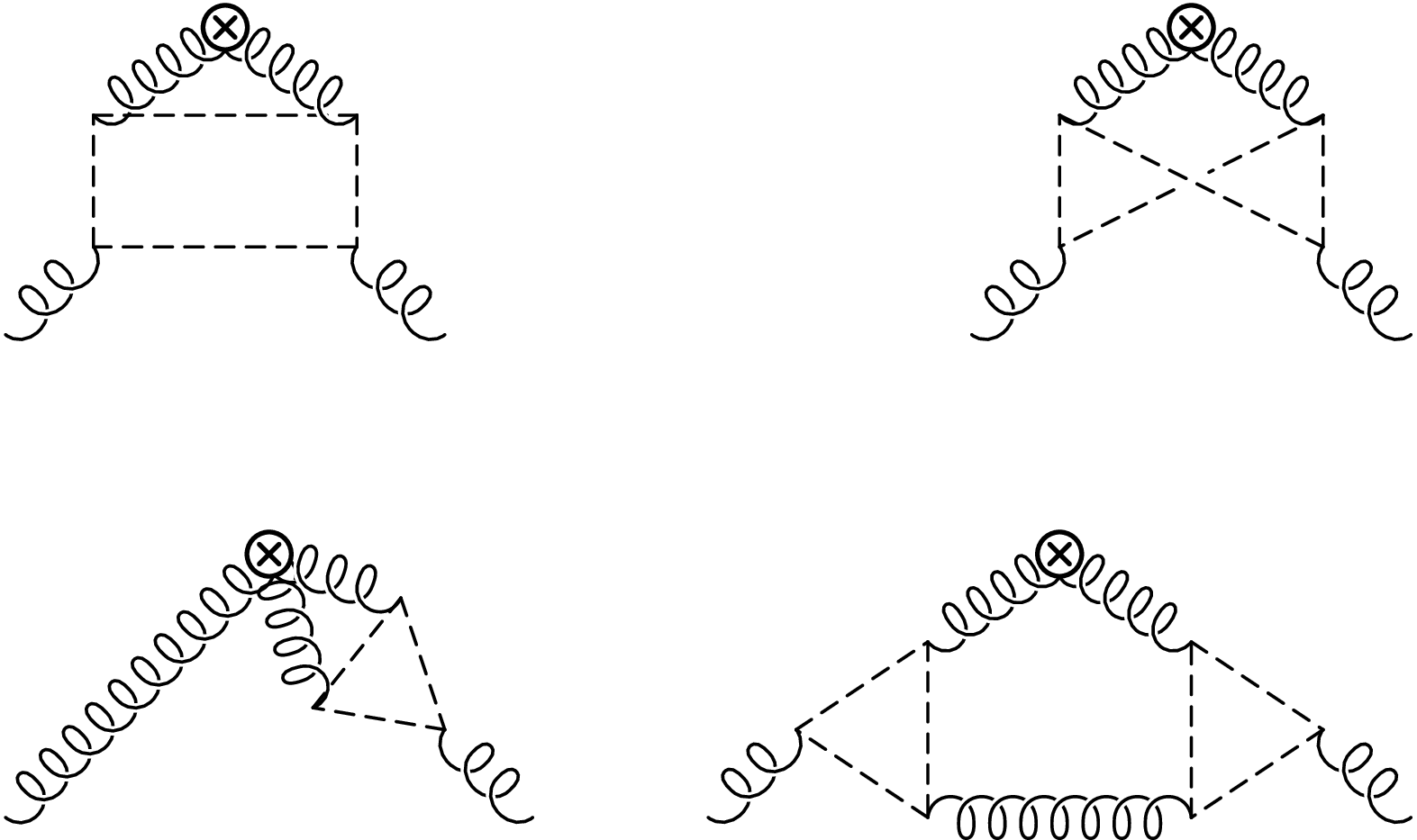}}

\vspace{0.75cm}
\hspace{3cm}{\bf Fig. 1. Leading order diagrams for $\lambda_+(a_c)$.}
\vspace{0.5cm}

\noindent 
For the unpolarised gluonic singlet operator we have,
\begin{eqnarray} 
\lambda_{+,1}(a_c) &=& 
-~ [8\mu^3n^2 + 8\mu^3n + 8\mu^3 + 2\mu^2n^4 + 4\mu^2n^3 - 22\mu^2n^2 
- 24\mu^2n \nonumber \\ 
&&~~~ - 28\mu^2 - 6\mu n^4 - 12\mu n^3 + 14\mu n^2 + 20\mu n + 32\mu 
+ 5n^4 \nonumber \\ 
&&~~~ + 10n^3 + n^2 - 4n - 12]\Gamma(n+2-\mu) 
\Gamma(\mu - 1)\mu C_2(R) \eta_1^{\mbox{o}} \nonumber \\ 
&&~~~~ /[(\mu - 2)^2(n + 2)(n + 1)(n - 1)
\Gamma(2 - \mu)\Gamma(\mu + n)n T(R)] \nonumber \\ && \nonumber \\  
&&+~ \frac{2\mu(\mu - 1)S_1(n) C_2(G) \eta_1^{\mbox{o}}} 
{(2\mu - 1)(\mu - 2)T(R)} \nonumber \\ 
&& \nonumber \\ 
&&-~ [32\mu^5n^2 + 32\mu^5n + 32\mu^5 - 144\mu^4n^2 - 144\mu^4n - 160\mu^4 
- 4\mu^3n^4 \nonumber \\ 
&&~~~ - 8\mu^3n^3 + 240\mu^3n^2 + 244\mu^3n + 316\mu^3 + 16\mu^2n^4 
+ 32\mu^2n^3 \nonumber \\ 
&&~~~ - 180\mu^2n^2 - 196\mu^2n - 306\mu^2 - 20\mu n^4 - 40\mu n^3 + 59\mu n^2
\nonumber \\
&&~~~ + 79\mu n + 146\mu + 8n^4 + 16n^3 - 6n^2 - 14n - 28]\mu C_2(G) 
\eta_1^{\mbox{o}} \nonumber \\ 
&&~~~~ /[8(2\mu - 1)(\mu - 1)^3(\mu - 2)(n + 2)(n + 1)(n - 1)n T(R)]  
\nonumber \\ && \nonumber \\  
&&+~ [32\mu^5n^2 + 32\mu^5n + 32\mu^5 + 8\mu^4n^4 + 16\mu^4n^3 - 120\mu^4n^2 
- 128\mu^4n \nonumber \\
&&~~~ - 160\mu^4 - 32\mu^3n^4 - 64\mu^3n^3 + 160\mu^3n^2 
+ 192\mu^3n + 316\mu^3 + 48\mu^2n^4 \nonumber \\ 
&&~~~ + 96\mu^2n^3 - 78\mu^2n^2 - 126\mu^2n - 306\mu^2 - 31\mu n^4 - 62\mu n^3 
+ 31\mu n \nonumber \\ 
&&~~~ + 146\mu + 7n^4 + 14n^3 + 7n^2 - 28]  
\Gamma(n+2-\mu)\Gamma(\mu - 1)\mu C_2(G) \eta_1^{\mbox{o}} \nonumber \\ 
&&~~~\, /[8(2\mu - 1)(\mu - 1)^2(\mu - 2)
(n + 2)(n + 1)(n - 1)\Gamma(2-\mu)\Gamma(\mu + n)n T(R)] \nonumber \\  
\label{answer} 
\end{eqnarray} 
implying,
\begin{eqnarray} 
d_{31} + \frac{b_{31} c_1}{d_{11}}  
&=& \frac{64(n^2 + n + 2)^2 (S_1(n))^2 C_2(R)}{3(n + 2)(n + 1)^2(n - 1)n^2 
T(R)} \nonumber \\ 
&& -~ \frac{64(10n^6 + 30n^5 + 109n^4 + 168n^3 + 155n^2 + 76n + 12) S_1(n) 
C_2(R)}{9(n + 2)(n + 1)^3(n - 1)n^3 T(R)} \nonumber \\  
&& -~ 4[33n^{10}\! + 165n^9 - 32n^8 - 1118n^7 - 5807n^6 \nonumber \\ 
&& ~~~~~ - 12815n^5 - 16762n^4 - 13800n^3 - 7112n^2 \nonumber \\
&& ~~~~~ - 2112n - 288]C_2(R)/[27(n + 2)(n + 1)^4(n - 1)n^4 T(R)] \nonumber \\ 
&& -~ \frac{8(8n^6 + 24n^5 - 19n^4 - 78n^3 - 253n^2 - 210n - 96) S_1(n) 
C_2(G)}{27(n + 2)(n + 1)^2(n - 1)n^2 T(R)} \nonumber \\ 
&& -~ 2[87n^8 + 348n^7 + 848n^6 + 1326n^5 + 2609n^4 + 3414n^3 + 2632n^2 
\nonumber \\ 
&& ~~~~~ + 1088n + 192]C_2(G)/[27(n + 2)(n + 1)^3(n - 1)n^3 T(R)]  
\label{dbcd} 
\end{eqnarray}
For the polarised gluonic singlet operator we have,
\begin{eqnarray} 
\lambda_{+,1}(a_c) &=& -~ \frac{(n + 2)(n - 1)\Gamma(n+2-\mu)\Gamma(\mu + 1) 
C_2(R)\eta_1^{\mbox{o}}}{(\mu - 2)^2(n + 1)\Gamma(2 - \mu)\Gamma(\mu + n)n 
T(R)} \nonumber \\ && \nonumber \\ && +~ \frac{2\mu(\mu - 1) S_1(n) C_2(G) 
\eta_1^{\mbox{o}}} {(2\mu - 1)(\mu - 2) T(R)} \nonumber \\ && \nonumber \\  
&& -~ [4\mu^3n^2 + 4\mu^3n - 8\mu^3 - 8\mu^2n^2 - 8\mu^2n + 16\mu^2 + 5\mu n^2 
\nonumber \\ 
&& ~~~ + 5\mu n - 9\mu - n^2 - n + 2] \Gamma(n+2-\mu) \Gamma(\mu)\mu C_2(G)  
\eta_1^{\mbox{o}} \nonumber \\ 
&& ~~~~~ /[8(2\mu - 1)(\mu - 1)^3(n + 1) \Gamma(3 - \mu)\Gamma(\mu + n)n T(R)] 
\nonumber \\ && \nonumber \\  
&& -~ [32\mu^4 - 4\mu^3n^2 - 4\mu^3n - 120\mu^3 + 16\mu^2n^2 + 16 \mu^2n 
\nonumber \\ 
&& ~~~ + 160\mu^2 - 20\mu n^2 - 20\mu n  - 89\mu + 8n^2 + 8n + 18] \mu C_2(G) 
\eta_1^{\mbox{o}} \nonumber \\ 
&& ~~~~~ /[8(2\mu - 1)(\mu - 1)^3(\mu - 2)(n + 1)n T(R)] 
\label{res} 
\end{eqnarray} 
giving,
\begin{eqnarray} 
d_{31} ~+~ \frac{b_{31}c_1}{d_{11}} &=& -~  
\frac{64(7n^2 + 7n + 3)(n + 2)(n - 1) S_1(n) C_2(R)}{9(n + 1)^3n^3} 
\nonumber \\ 
&& +~ \frac{64(n + 2)(n - 1) S_1^2(n) C_2(R)}{3(n + 1)^2n^2} \nonumber \\ 
&& -~ 4[33n^8 + 132n^7 + 142n^6 - 36n^5 - 263n^4 - 312n^3 \nonumber \\ 
&& ~~~~ + 280n^2 + 408n + 144] C_2(R)/[27(n + 1)^4n^4] \nonumber \\ 
&& -~ \frac{8(8n^4 + 16n^3 - 19n^2 - 27n + 48) S_1(n) C_2(G)}{27(n + 1)^2n^2}  
\nonumber \\ 
&& -~ \frac{2(87n^6 + 261n^5 + 249n^4 + 63n^3 - 76n^2 - 64n - 96) C_2(G)} 
{27(n + 1)^3n^3}  
\label{3loop} 
\end{eqnarray} 
Throughout these calculations we used {\sc reduce} \cite{23} and {\sc form} 
\cite{24} to handle tedious amounts of algebra. The quantity 
$\eta_1^{\mbox{o}}$ is defined by 
\begin{equation} 
\eta_1^{\mbox{o}} ~=~ \frac{(2\mu-1)(\mu-2)\Gamma(2\mu)} 
{4\Gamma^2(\mu)\Gamma(\mu+1)\Gamma(2-\mu)} 
\end{equation}  
and we have set $\lambda_+(a_c)$ $=$ $\sum_{i=0}^\infty 
\lambda_{+,i}(a_c)/\Nf^i$. The finite sum $S_1(n)$ is given by $S_1(n)$
$=$ $\sum_{i=1}^n1/i$.

These expressions agree exactly with all known perturbative results. This can 
be seen by putting $\mu$ $=$ $2$ $-$ $\epsilon$ in (\ref{dbcd}) and 
(\ref{3loop}) and expanding in powers of $\epsilon/\Nf$. An interesting feature
of the mixing matrix (\ref{mixmat}) is that the $O(1/\Nf)$ contribution to 
$\gamma_{gq}(a)$ depends only on the Casimir $C_2(R)$. This is evident from the
one and two loop results for all $n$ and the three loop results for $n$ $\leq$ 
$8$. Assuming this to be true for all $n$ at three loops we can see from 
(\ref{dbcd}) and (\ref{3loop}) that we can deduce the exact form of the 
coefficients $d_{31}$ in the $C_2(G)$ sector. This means that by using an 
inverse Mellin transform we can calculate the $O(1/\Nf)$ part of the three loop
DGLAP function $P_{gg}$ for both unpolarised and polarised cases. For the 
unpolarised splitting function we obtain,
\begin{eqnarray}
P_{gg}(d_{31},C_2(G)) &=& \frac{1}{4} C_2(G) \left( ~\frac{64}{27} 
\left[\frac{1}{1-x} \right]_+ -~ \frac{64}{27} ~-~ \frac{58}{9}\delta(1-x) ~+~ 
\frac{128}{9}(x+1)\mbox{Li}_2(x) \right. \nonumber \\ 
&& \left. +~ \frac{8}{27}~ \frac{(x-1)(52x^2+19x+52)}{x}\ln(1-x) ~-~ 
\frac{128}{9}\psi^\prime(1)(x+1)\right. \nonumber \\
&& \left. -~  \frac{8}{27}~(52x^2+43x+76)\ln(x) ~+~ \frac{32}{9}(x+1)\ln^2(x)
\right. \nonumber \\
&& \left. +~ \frac{8}{81}\frac{(x-1)(236x^2+47x+236)}{x} \right) 
\end{eqnarray}
Similarly for the polarised splitting function,
\begin{eqnarray}
P_{gg}(d_{31},C_2(G)) &=& \frac{1}{4} C_2(G) \left( ~\frac{64}{27} 
\left[\frac{1}{1-x} \right]_+ -~ \frac{64}{27} ~-~ \frac{58}{9}\delta(1-x) ~+~ 
\frac{128}{9}(x+1)\mbox{Li}_2(x) \right. \nonumber \\ 
&& \left.   -~ \frac{128}{9}\psi^\prime(1)(x+1) ~+~ \frac{32}{9}(x+1)\ln^2(x)
{}~-~ \frac{8}{27}(67x-56)\ln(x) \right. \nonumber \\
&& \left. -~ \frac{328}{9}(1-x)\ln(1-x)
{}~-~ \frac{920}{27}(x-1) \right)
\end{eqnarray}

These results complete the programme to calculate the $O(1/\Nf)$ corrections 
for the anomalous dimensions of the twist-2 light cone operators. At present it 
seems that a continuation of this programme to include contributions of 
$O(1/\Nf^2)$ may be viable. We conclude by noting that the results may also be 
useful in estimating the full three loop corrections to the operator dimensions
by using asymptotic Pad\'{e} approximant techniques \cite{25}.

{\bf Acknowledgements.}  
The author would like to thank the hosts of QUARKS '98 for their extremely kind
hospitality and is also grateful to many members of the workshop for helpful 
and interesting discussions. This work was carried out with the support of
{\sc PPARC} through an Advanced Fellowship (JAG) and a Postgraduate
Studentship (JFB).

\end{document}